\documentclass[aps,pre,twocolumn,reprint,longbibliography,titlepage,showpacs,showkeys,amsfonts,amssymb,amsmath]{revtex4-1}
\usepackage{bm}
\usepackage{pdfsync}
\usepackage{color}
\newcommand{\bse}{\begin{subequations}}
\newcommand{\ese}{\end{subequations}}
\newcommand{\be}{\begin{equation}}
\newcommand{\ee}{\end{equation}}
\newcommand{\bea}{\begin{eqnarray}}
\newcommand{\eea}{\end{eqnarray}}
\newcommand{\kb}{k_{_{\mathrm{B}}}}
\newcommand{\phat}{\widehat{\mathrm{p}}}
\newcommand{\qhat}{\widehat{q}}
\newcommand{\rhohat}{\widehat{\rho}}
\newcommand{\tp}{t^{\prime}}
\newcommand{\phiq}{\varphi_{_{\!\mathbf{q}}}}
\newcommand{\phiv}{\varphi_{_{\!\mathbf{v}}}}
\newcommand{\chiq}{\chi_{_{\mathbf{q}}}\!}
\newcommand{\chiv}{\chi_{_{\mathbf{v}}}\!}
\newcommand{\sigmaQ}{\sigma_{\mbox{\tiny{\!\bf{Q}}}}}
\newcommand{\DQ}{D_{_{\mbox{\tiny{\!\bf{Q}}}}}\!}
\newcommand{\sigmaone}{\sigma_{_{\!\mathbf{1}}}}
\newcommand{\Done}{D_{_{\mbox{\tiny{\!\bf{1}}}}}\!}
\newcommand{\pcla}{p_{_{\mbox{\tiny{\!\bf{CL}}}}}}
\newcommand{\Dclas}{D_{_{\mbox{\tiny{\!\bf{CL}}}}}\!}
\newcommand{\sigmac}{\sigma_{_{\mbox{\tiny{\!\bf{CL}}}}}}
\newcommand{\qzero}{q_{_{\mathbf{0}}}}
\newcommand{\vzero}{v_{_{\mathbf{0}}}}
\newcommand{\nun}{\nu_{_{\!\mathrm{n}}}}
\newcommand{\lambdaone}{\lambda_{_{\mathbf{1}}}}
\newcommand{\lambdatwo}{\lambda_{_{\mathbf{2}}}}
\allowdisplaybreaks

\begin{document}
\title{Fokker-Planck equation of the reduced Wigner function  associated to an Ohmic  quantum Langevin dynamics.}
\author{Pedro J. Colmenares}
\thanks{Corresponding Author}
\email{colmenar@ula.ve, gochocol@gmail.com}
\affiliation{Departamento de Qu\'{\i}mica, Universidad de Los Andes.
 M\'erida 5101, Venezuela}
%\date{\today}

%-------------------
%-------------------
\begin{abstract}
This article has to do with the derivation and solution of the Fokker--Planck equation associated to the  momentum-integrated Wigner function of a particle subjected to a harmonic external field in contact with a ohmic thermal bath of quantum harmonic oscillators. The strategy employed is a simplified version of the phenomenological approach of Schramm, Jung and Grabert of interpreting the operators as c numbers to derive the  quantum master equation arising from a twofold transformation of the Wigner function of the entire phase space. The statistical properties of the random noise comes from the integral functional theory of Grabert, Schramm and Ingold. By means of a single Wigner transformation, a simpler equation than that mentioned before is found. The Wigner function reproduces the known results of the classical limit. This allowed us to rewrite the underdamped classical Langevin equation as a first-order stochastic differential equation with time-dependent drift and diffusion terms.\end{abstract}
%-------------------
%-------------------

 \pacs{05.30.Ch; 05.30.?d; 05.40.Jc; 02.50.Ey}
\keywords{Quantum statistical mechanics, Brownian motion, Stochastic processes, Fokker Planck equation.}
\maketitle

%---------------------------
%---------------------------
\section{Introduction.}
%---------------------------
%---------------------------
For a particle trapped in a harmonic potential, for instance, in an optical trap, immersed in a thermal bath, various  approaches can be used to describe its dynamics. Let the Langevin equation be the chosen one. Now, it is asked, what is the probability the particle actually be at $q$? Classically, this question has been already answered in the calculation of the maximum work performed by the particle, when the system is driven by a particular measurement protocol. This has been done as much for the Markovian underdamped \cite{SeifertAbreu} and overdamped\cite{GomezSchmiedlSeifert} and, in the generalized  regime\cite{OscarErnestoPedro}, as well. But, what about if the thermal reservoir behaves quantumly? This is the aim of the article: starting from the Ohmic version of the quantum Langevin equation (QLE), the reduced Wigner function $W(q,t )$, which gives the desired probability density, will be derived from the solution of its associate Fokker-Planck equation (FPE).

The first derivation of the quantum gen\-er\-al\-ized Lan\-ge\-vin equation (QGLE) dated back to 1965 in the work by Ford {\it et al.} \cite{FordKacMazur}. They consider a set of $2N+1$ interacting harmonic oscillators and focus their work on studying the dynamics in a particular one due to the effect of the rest of oscillators acting as a heat bath. The resulting operators equation resembles the classical generalized Langevin equation in coordinate space. 
Another derivation is that of Ford {\it et al.} \cite{FordLewisOConnell}. Starting from the Heisenberg equation of motion,  they provided a purely quantum derivation of the QGLE with a random force operator acting on the Hilbert space of the entire system. 
Limiting the analysis  to  Ohmic baths, i.e. where the hydrodynamics drag depends on the instantaneous velocity,  Ford and Kac\cite{FordKac} found out that the QLE should have an  extra term, the so called ''slip term'', which has an infinitesimal dependence in time. It vanishes after a very short period of relaxation. In the strict Ohmic dissipation, it reduces to a $\delta$-contribution. As H\"{a}nggi\cite{Hanggi2} indicated, it is frequently omitted for $t>t_{0}$ although it will affect the trajectory of the phase point. For the purposes of physical applications, this fine detail can be put aside, but not mathematically, since the universality of the Langevin equation would not be guaranteed in the quantum regime\cite{FordKac}. By universal it is meant that many physical problems satisfy an equation whose form shall be equal for all. 
An important result to assure the validity of the Ohmic
 QLE was the finding of Benguria and Kac\cite{BenguriaKac} about the requirement that to get a quantum 
 mechanical canonical  probability distribution, the noise operator has to be a purely Gaussian 
 process. It is important to remark that 
 van Kampen\cite{VanKampen3} proved that the QLE is true only in the lowest level of interaction 
 between the particle and the thermostat, that is, when the interaction is bilinear. In all these works, the noise operator acts over the Hilbert space of the entire system.
This brings as consequence that the bath coordinates cannot be eliminated and the utility of both, the QGLE and QLE, is of limited practical usefulness\cite{Hanggi2}.  A QGLE derived from the Heisenberg equation of motion acting  solely on the Hilbert space of the system simply does not exist\cite{Hanggi2}. Even worse, badly managed stochastic approximations on the colored  noise and time-dependent friction, have a definite impact in the breaking of the quantum behavior. ``{\it Where one could go wrong}''and many other fundamental aspects about the QGLE can be found in the review by H\"{a}nggi and Ingold\cite{HanggiIngold}.
 
The route chosen to successfully eliminate the undesirable bath coordinates was throughout path integrals of the Feynman-Vernon theory of damped quantum systems\cite{FeynmanVernon}. Thus, Caldeira and Leggett\cite{CaldeiraLeggett} applied such a theory to study the quantum  dissipation without the presence of initial correlations between the particle and the bath, that is, without the dependence of the initial preparation over the evolution of the system. They found a semiclassical Fokker Planck equation by making a transformation from Hilbert space to the classical phase space by way of the Wigner distribution functions. 
Grabert {\it et al.} \cite{GrabertSchrammIngold} refined the Feynman-Vernon approach and developed the functional integral method which include the initial preparation. It provides an exact description of the system in terms of the mass of the particle, the spectral density of the bath and the external potential. Although the equations are rather complex, they reduce to manageable mathematical objects for  Ohmic baths. They were used by Schramm, Jung and Grabert (SJG) \cite{SchrammJungGrabert} in the exact derivation of the Wigner function using a phenomenological approach based on considering the operators as c numbers in the whole phase space. In the classical limit, it agrees with the findings by Adelman\cite{Adelman1}. Additionally, they derived the master equation for the quantum operators  (QME), also known by some authors \cite{Gardiner2} as the adjoint equation, using the transformation rules from Wigner function to quantum operators \cite{Gardiner3,ArnoldFagnolaNeumann}. Later, Karrlein and Grabert \cite{KarrleinGrabert}, validate these findings by demonstrating that the functional integral approach for an initial thermal preparation function, reproduces the generalized FPE for the classical harmonic oscillator of Adelman\cite{Adelman1}. They also found that the Wigner function is that of SJG and furthermore show that an  initial factorization of the entire density matrix does not yield the Adelman equation. 
In 2006, Isar and Sandulescu\cite{IsarSandulescu} derived among other things, the FPE for the Wigner function. They show that the Wigner quasiprobability distribution
is a two-dimensional Gaussian with a width determined by the diffusion coefficients.
With a theory of their own, Ding {\it et al.} \cite{DingEtAl} recently review Caldeira and Leggett theory to develop a  QME by approximating the bath coordinates correlation function with a bi-exponential function. For its own nature, this approximated work is out and beyond the scope of the goals of this research. 

The purpose of the present work is to rephrase the exact derivation procedure of SJG such that it simplifies the structure of the QME of those already mentioned. As a subproduct, it is submitted  as a first order stochastic differential equation (SDE) complementary to the classical Langevin equation. 

The manuscript is composed of two main parts. The general theory is developed in the first with some auto explanatory sections. A Concluding Remarks section closes the article, and an Appendix is included to complement some derivations.

%------------------------------------------
%------------------------------------------
\section{General Theory.}
%------------------------------------------
%------------------------------------------
This section parcels the main results of the proposed method. The derivation and solution of the Fokker-Planck equation (FPE) associated to the reduced Wigner function of the system is presented in a first place. The QME is subsequently derived  in the second section and thirdly, the corresponding classical limit is investigated comparing the results achieved with those already known. Finally, in the fourth part,  a first order SDE equivalent to the classical Markovian Langevin equation is proposed.
%-------------------------------------------
%-------------------------------------------
\subsection{The FPE for the reduced Wigner function.}
%------------------------------------------
%------------------------------------------
Subscribing entirely the phenomenological approach of SJG\cite{SchrammJungGrabert} of considering the QLE as a c-number equation, the dynamics of a particle with mass $M$, subjected to an external potential $\omega_{0}^{2}q(t)/2$, in contact with an Ohmic bath of quantum harmonic oscillators at  temperature $T$ with a friction coefficient $\gamma$ is given by:
\be
\ddot{q}(t)=-\gamma\,\dot{q}(t)-\frac{\omega_{0}^{2}}{M}\,q(t)+\frac{1}{M}\,\xi(t),\\
\label{QLE}
\ee
where $\xi(t)$ is the Gaussian quantum noise with zero mean and two--time correlation function\cite{GrabertSchrammIngold,SchrammJungGrabert}:
\bea
\left<\xi(t)\,\xi(s)\right>=&-&\left(\frac{\gamma M}{2\beta}\right)\nu\sinh\left[\frac{1}{2}\nu(t-s)\right]^{-2}\nonumber\\
&+&\,i\,\gamma\, M\,\hbar\,\dot{\delta}(t-s).\label{corxi}
\eea
Here, $\hbar$ is the Planck constant divided by $2\pi$, $\beta=(\kb T)^{-1}$, $\kb$ is the Boltzmann constant, and frequency $\nu=2\pi(\hbar\beta)^{-1}$. Unlike the Markovian Langevin equation, the noise has a colored spectrum and is correlated with $q(0)$ due to the preparation procedure. The latter is \cite{SchrammJungGrabert}:
\be
\left<\xi(t)q(0)\right>=-\frac{2\,\gamma}{\beta}\sum_{\mathbf{n}=1}^{\infty}\frac{\nun}{(\nun\!+\!\lambdaone)(\nun\!+\!\lambdatwo)}\mbox{e}^{-\nun t},\label{initcorr}\\
\ee
where $\nun=n\nu$ is the Matsubara frequency and frequency  $ \lambda_{_{\mathbf{1,2}}}=[\gamma\pm(\gamma^{2}- 4\omega_{0}^{2}/M)^{1/2}\,]/2$. The sum can be solved to give \cite{mathematica}
\begin{widetext}
\be
\left<\xi(t)q(0)\right>=-\frac{2\,\gamma}{\beta}\frac{\mbox{e}^{-\nu t}}{(\lambdaone-\lambdatwo)(\lambdaone+\nu)(\lambdatwo+\nu)}%\nonumber\\
\Bigg[\lambdaone(\lambdatwo+\nu)\,\,{}_{2} F_{1}\big(1;a_{_{\mathbf{1}}};b_{_{\mathbf{1}}};\mbox{e}^{-\nu t}\big)%\nonumber\\
-\lambdatwo(\lambdaone+\nu)\,\,{}_2 F_1\big(1;a_{_{\mathbf{2}}};b_{_{\mathbf{2}}};\mbox{e}^{-\nu t}\big)\Bigg],\label{corq0xi}
\ee
\end{widetext}
where the parameters  $a_{_{\mathbf{1,2}}}=\big(\lambda_{_{\mathbf{1,2}}}+\nu\big)/\nu$ and $b_{_{\mathbf{1,2}}}=2+\lambda_{_{\mathbf{1,2}}}/\nu$ are two of the arguments of the  hypergeometric series
 ${}_{2}F_{1}(A;B;C;x)=\sum_{n=0}^{\infty}[(A)_{n}(B)_{n}/(C)_{n}]x^{n}/n!$, respectively\cite{Abramowitz}. 

Applying the Laplace transform to Eq. (\ref{QLE}) and inverting, the solution of the QLE and its derivative read as:
\bse\bea
q(t)&=&\chiq(t)q(0)+\chiv(t)v(0)+\phiq(t),
\label{qt}\\
\dot{q}(t)&=&\dot{\chiq}(t)q(0)+\dot{\chiv}(t)v(0)+\phiv(t),\label{qprime}
\eea\ese
where the noises $\phiq(t)$ and $\phiv(t)$  are functional of $\xi(t)$ and the Laplace transforms of the susceptibilities $\chiq(t)$ and $\chiv(t)$  are, respectively:
\bse
\bea
\phiq(t)&=&\frac{1}{M}\int_{0}^{t}d\tp\chiv(t-\tp)\xi(\tp),\label{varphiq}\\
\phiv(t)&=&\frac{1}{M}\int_{0}^{t}d\tp\dot{\chiv}(t-\tp)\xi(\tp),\label{varphiv}\\
\chiq(s)&=&(s+\gamma)\chiv(s),\label{chiqs}\\
\chiv(s)&=&\frac{M}{Ms(s+\gamma)+\omega_{0}^{2}}.\label{chivs}
\eea
\ese
A relationship between the susceptibilities is obtained from Eq. (\ref{chiqs}), i.e.,  $\chiq(t)=1-\omega_{0}^{2}M^{-1}\!\!\int_{0}^{t}d\tp\chiv(\tp)$ \cite{AdelmanGarrison,WorPJPhysA}. Their explicit dependence on $\gamma$ and $\omega_{0}$ is shown in Sec. \ref{limiteclasico}.

For a given realization of the quantum noise, Eq. (\ref{qprime}) describes a flow in $q$ space. The density of this flow, $f(q[\xi(t)],t)$, where the functional dependence of the coordinate on the noise has been made explicitly, evolves in time according the continuity equation:
\be
\frac{\partial f(q[\xi(t)],t)}{\partial t}=-\frac{\partial}{\partial q}\Bigg[\dot{q}[\xi(t)]f(q([\xi(t)],t)\Bigg]\label{continuity}
\ee
For an ensemble of trajectories, the probability density is just the ensemble average of  $f(q[\xi(t)],t)$ \cite{VanKampen2}, i.e., 
\bea
p(q,t,\qzero,\vzero)&=&\Big<f(q[\xi(t)],t\Big>_{\!\!\xi},\nonumber\\
&=&\Big<\delta(q(t)-q)\delta(q(0)-\qzero)\Big>_{\!\!\xi},\nonumber
\eea
where the subindex $\xi$ indicates that the average has to be taken over the distribution of the quantum noise. Then, carrying out the proper substitutions in Eq. (\ref{continuity}), 
\bea
\frac{\partial p(q,t,\qzero,\vzero)}{\partial t}=&-&\frac{\partial}{\partial q}\Big<\phiv(t)\delta(q(t)-q)\delta(q(0)-\qzero)\Big>_{\!\xi}\nonumber\\
&-&\overline{v}(t)\frac{\partial p}{\partial q},\label{Eq2}
\eea
where for short $\overline{v}(t)=\dot{\chiq}(t)\qzero+\dot{\chiv}(t)\vzero$ is the drift velocity. 

As in SJG\cite{SchrammJungGrabert},  expanding the average of Eq. (\ref{Eq2})  in terms of the cumulants of the noise and, adding upon the extra term arising from the correlation between $\xi(t)$ and $q(0)$, the following is obtained:
\begin{widetext}
\be
\frac{\partial p(q,t,\qzero,\vzero)}{\partial t}
=-\overline{v}(t)\frac{\partial p}{\partial q}+\frac{\partial}{\partial q}\Bigg[\int_{0}^{t}d\tp\Big<\phiv(t)\phiv(\tp)\Big>\frac{\partial p}{\partial q}-\Big<\xi(t)q(0)\Big>\frac{\partial }{\partial q(0)}\Big<\delta(q(t)-q)\delta(q(0)-\qzero)\Big>_{\!\!\xi}\Bigg].\label{eq8}
\ee
\end{widetext}
Knowing that $\partial\delta(x-y)/\partial x =-\partial\delta(x-y)/\partial y$, then:
\be
\frac{\partial }{\partial q(0)}\Big<\delta(q(t)-q)\delta(q(0)-\qzero)\Big>_{\!\!\xi}=-\frac{\partial p}{\partial\qzero},
\label{eq9}
\ee
and from Eq. (\ref{qt}) $\partial /\partial\qzero =\chiq(t)\partial /\partial q$ for a fixed value of  $\qzero$. Then the final equation for the evolution of the probability density $p(q,t,\qzero,\vzero)$ is:
 \bea
\frac{\partial p(q,t,\qzero,\vzero)}{\partial t}=&-&\overline{v}(t)\frac{\partial p}{\partial q}+\frac{1}{2}\Done(t)\frac{\partial^{2}p}{\partial q^{2}},\label{QFPE}
\eea
where the function $\Done(t)$ is defined as
\bea
\Done(t)=&2&\Bigg[\!\int_{0}^{t}d\tp\Big<\phiv(t)\phiv(\tp)\Big>
+\chiq(t)\Big<\xi(t)q(0)\Big>\Bigg].\label{D1clas}
\eea

The linear transformations $y=q-\int_{0}^{t}d\tp\overline{v}(\tp)$ and $r=\int_{0}^{t}d\tp D_{_{\mbox{\tiny{\!Q}}}}\!(\tp)$ applied to Eq. (\ref{QFPE}) give the simple diffusion equation \cite{Ratcliff}:
\be
\frac{\partial p}{\partial r} = \frac{1}{2}\frac{\partial^{2}\,p}{\partial \,y^{2}},\label{simple}
\ee
whose solution for the initial condition $p(y,0,y_{_{\mathbf{0}}},\vzero)=\delta(y-y_{_{\mathbf{0}}})$ 
is a Gaussian centered at $y_{_{\mathbf{0}}}$ with standard deviation $r$. Transforming to the original variables gives:
\bse\bea
p(q,t|\qzero,\vzero)&=&\!\frac{1}{\sqrt{2\pi\,\sigmaone\!(t)}}
\exp\left[-\frac{(q-\overline{q}(t))^{2}}{2\,\sigmaone\!(t)}\right],\\
\overline{q}(t)&=&\chiq(t)\,\qzero+\chiv(t)\,\vzero,\\
\sigmaone\!(t)&=&\int_{0}^{t}\!\!d\tp \Done(\tp)\label{A11}.
\eea\ese

In general, position and velocity relax to equilibrium at different rates. The velocity achieves the canonical distribution faster than position\cite{WorPJPhysA}. Then, the density $p(q,t\,|\, \qzero)$ given an initial thermal distribution of initial velocity $\vzero$ is found by averaging the solution $p(q,t\,|\, \qzero,\vzero)$ over the Maxwell velocity distribution. The result  for an initial thermal condition will be:
\bse\bea
p(q,t | \qzero)&\!=\!&\!\frac{1}{\sqrt{2\pi\sigmaQ(t)}}
\!\exp\left[-\frac{(q-\chiq(t)\qzero)^{2}}{2\,\sigmaQ(t)}\right],\label{PJprob1}\\
\sigmaQ(t)&=&\sigmaone\!(t)+\frac{\kb T}{M}\chiv^{\!\!2}(t).\label{SigmaAveraged}
\eea\ese

The FPE given as a solution of Eq. (\ref{PJprob1}) is obtained by a procedure due to Adelman and Garrison (AG) \cite{AdelmanGarrison} and shown in the Appendix. It gives:
\bse\bea
\frac{\partial p(q,t,\qzero)}{\partial t} &=&-\Omega(t)
\frac{\partial}{\partial q}\Big[q\,p\Big]+\frac{1}{2}\DQ(t)\frac{\partial^{2}p}{\partial q^{2}},\label{adelman1}\\
\Omega(t)&=&\frac{\dot{\chiq}(t)}{\chiq(t)},\label{Omega1}\\
\DQ(t)&=&\dot{\sigmaQ}(t)-2\,\sigmaQ(t)\,\Omega(t).\label{Dadelman}
 \eea\ese

Since the reduced Wigner function $W(q,t)$ obeys the relations\cite{Feynman}:
\bea
W(q,t)&=&p(q,t|\qzero)=\frac{1}{2\pi\hbar}\int_{-\infty}^{\infty}dp\,W(q,p,t),\label{feynman}
\eea
the FPE associated to $W(q,t)$  has the same form of Eq. (\ref{adelman1}). The desired result is then:
\be
\frac{\partial W(q,t)}{\partial t} =-\Omega(t)\frac{\partial }{\partial q}\big[q\,W\big]
+\!\frac{1}{2}\DQ(t)\frac{\partial^{2}W}{\partial q^{2}},\label{FPEWigner}
\ee
whose solution is identical to Eq. (\ref{PJprob1}). The density matrix $\rho(q,t)$ is also the reduced Wigner function\cite{Feynman}.

It is important to remark that there are other methods  to find the reduced Wigner function besides that developed here. For instance, one can appeal to Eq. (\ref{feynman}) to get it by integrating SJG's Wigner function in the momentum space. Even more, the procedure developed by Ford and O'Connell\cite{FordOConnell} can also be used to find the solution of the quantum master equation of Hu {\it et al.} cite{HuPazZhang}.  In any case, whatever procedure is employed, all of them should give the same answer. In particular, by including in the master equation of Hu {\it et al.}, the dependence of the initial correlation between the system and the thermal bath over the dynamics of the Brownian particle, two results will show up. On one hand, Hu {\it et al.} should replicate the QME of SJG and second, the procedure developed in this work will agree with that of Ford and O'Connell. The main difference between the last two approaches is just operational. 

%-------------------------------------------
%-------------------------------------------
\subsection{The quantum master equation.}
%------------------------------------------
%------------------------------------------
In quantum mechanics, the QME can be obtained using the transformation rules of the Wigner function in terms of the density operator $\rhohat(t)$\cite{Gardiner3,ArnoldFagnolaNeumann,Gardiner2}. The rules for $W(q,t)$ and $W(q,p,t)$ are the same, so the two partial derivatives in Eq. (\ref{FPEWigner})  are transformed as:
\bea
\frac{\partial }{\partial q}\big(qW\big)=W+q\frac{\partial W}{\partial q}&\longrightarrow&\rhohat+\frac{i}{2\hbar}\Big\{\qhat,\Big[\phat,\rhohat\Big]\Big\},\\
\frac{\partial^{2}W}{\partial q^{2}}&\longrightarrow&-\frac{1}{\hbar^{2}}\Big[\phat,\Big[\phat,\rhohat\Big]\Big],
\eea
where $\phat=-i\hbar\partial/\partial q$\ is the momentum operator and $\{\widehat{A},\widehat{B}\}=(\widehat{A}\cdot \widehat{B}+\widehat{B}\cdot \widehat{A})$ denotes the anticommutator of $\widehat{A}$ and $\widehat{B}$.

The reduced QME will be:
\bea
\frac{\partial\rhohat}{\partial t}=&-&\Omega(t)\Bigg(\rhohat+\frac{i}{2\hbar}\Big\{\qhat,\Big[\phat,\rhohat\Big]\Big\}\!\!\Bigg)\nonumber\\
&-&\frac{1}{2\hbar^{2}}\DQ(t)
\Big[\phat,\!\Big[\phat,\rhohat\Big]\Big].\label{adjointpj}
\eea
Compare this result with Eq. (38) of SJG\cite{SchrammJungGrabert} derived from the Wigner distribution of the whole phase space: \begin{widetext}
\bea
\frac{\partial\rhohat}{\partial t}&=&\widetilde{\gamma}(t)\rhohat-\frac{i}{M\hbar}\phat\left[\phat,\rhohat\right]
-\frac{i}{\hbar}M\tilde{\omega}^{2}(t)\qhat\Big[\qhat,\rhohat\Big]%\nonumber\\
-\frac{i\widetilde{\gamma}(t)}{\hbar}\phat\Big[\qhat,\rhohat\Big]
-\frac{i}{2M\hbar}\Big[\phat,\Big[\phat,\rhohat\Big]\Big]\nonumber\\
&-&\left\{\frac{iM\widetilde{\omega}^{2}(t)}{2\hbar}+\frac{\widetilde{\gamma}(t)\left<\mathrm{p}^{2}\right>}{\hbar^{2}}\right\}\Big[\qhat,\Big[\qhat,\rhohat\Big]\Big]%\nonumber\\
+\!\!\left\{\frac{M\widetilde{\omega}^{2}(t)\left<q^{2}\right>}{\hbar^{2}}-\frac{\left<\mathrm{p}^{2}\right>}{M\hbar^{2}}-\frac{i\widetilde{\gamma}(t)}{2\hbar}\right\}\!\!\Big[\phat,\Big[\qhat,\rhohat\Big]\Big],\label{SJG}
\eea
\end{widetext}
where the coefficients $\widetilde{\gamma}(t)$ and $\widetilde{\omega}(t)$ are defined in the text\cite{SchrammJungGrabert}. They are complex and become real in the classical limit \cite{KarrleinGrabert}. The averages $\left<q^{2}\right> $ and $\left<\mathrm{p}^{2}\right>$ are defined in Ref. \cite{GrabertWeissTalkner}. 
These two results are exact because no approximations were invoked in obtaining them, the former being structurally simpler than the latter. 

Now, although these two equations describe the same phenomenon from different approaches, they seem to generate rather different outcomes for the same set of initial conditions. It may be that, in explicit calculations, both equations will render the same results whenever the density operator is used to determine the properties of the system. Such a calculation is beyond the scope of the present work. However, as expected, since they share a common classical origin, it will be shown in the next section that the probability density of the present work coincides with the SJG's density integrated over the momentum space.

Finally, although it is not one of the main objectives of this research, it is important to make some remarks about the mathematical proof of  the quantum equivalence between Eqs. (\ref{adjointpj}) and (\ref{SJG}).

First of all, some thing has to be said about the methodology employed to include the quantum correlations  $\big<\xi(t)q(0)\big>$ and $\big<\phiv(t)\phiv(\tp)\big>$ in the two descriptions. In this work, the transformation given by Eq. (\ref{eq9}) allows to write the diffusive term $\DQ(t)$ of Eq. (\ref{adjointpj}) by preserving the structure of the correlations in the way that they were derived. A different strategy was used by SJG. Using their Eq. (32),  the effect of the correlations is absorbed in the functions $\widetilde{\omega}(t)$ and $\widetilde{\gamma}(t)$, which are linear combinations of both the position correlation functions and of its first, second, and third time derivative. The algebra to go from one equation to the other is cumbersome so, that another scheme could be invoked to get an insight about such an equivalence. That is, in the absence of a general proof, then it would be appropriate instead, to find a correspondence between $\Omega(t)$ with $\widetilde{\gamma}(t)$ and $\DQ(t)$ with $\widetilde{\omega}(t)$\footnote{Thanks are extended to one of the referees for drawing attention to this detail.}. Applying this strategy, whose algebra is also relatively lengthy, it would allow us to find a way to get a partial compatibility between Eqs. (\ref{adjointpj}) and (\ref{SJG}).

There is a fact for sure, independently of the approach, that any average calculation  should have to give the same result. This was done previously by Grabert {\it et al.} \cite{GrabertWeissTalkner} in the derivation of the different correlations  involving position and momentum for the c-number Ohmic QLE. Obviously, these results do not prove that Eqs. (\ref{adjointpj}) and (\ref{SJG}) are equivalent in the quantum regime, but they give a glimpse that there must be a proper way to prove it.

For the sake of simplicity, let $\rho(p,q,t)$ be the SJG's matrix density. For  any observable $\mathcal{O}$, its dispersion is given by $<\mathcal{O}^2\big>= \mbox{Tr}\big\{\widehat{\mathcal{O}}^{2}\widehat{\rho}\big\}=\int dp\int dx\, \mathcal{O}^{2}\rho(x,p,t)$. However according to Eq. (\ref{feynman}), this is equivalent to $<\mathcal{O}^2\big>=\int dx\,\mathcal{O}^{2}\rho(x,t)$. That is, the statistical average coincides whatever scheme is invoked, which means that the density matrix obtained from any of the operator equations will provide the same statistical result. The main difference is that the QME of this work is simpler that of SJG.

The above is a consequence of the contraction applied in this proposal. Since the rules to transform the Wigner function dynamics into one involving the density operator are the same in the two procedures, the alluded  contraction should be kept in the quantum regime. It is transferred to the functions $\Omega(t)$ and $\DQ(t)$ but not to the density operator. The physical interpretation of the latter is unique and has to be independent of its derivation. In this sense, the previous argument permits us to circumvent any algebraic manipulation needed to prove the equivalence between the Eqs. (\ref{adjointpj}) and (\ref{SJG}). They will provide same statistical results for any operator properly defined in the Hilbert space.

%-----------------------------------------------------
%-----------------------------------------------------
\subsection{The classical limit.}
\label{limiteclasico}
%-----------------------------------------------------
%-----------------------------------------------------
It corresponds to take $\hbar\rightarrow 0$. In this regime, the correlation
between the noise $\xi(t)$ and $q(0)$ vanishes,  $\big<\xi(t)q(0)\big>=0$,  and the two-time noise correlation reduces to the well-known Markovian expression $\big<\xi(t)\xi(\tp)\big>=2\kb T \gamma M\delta(t-\tp)$\cite{SchrammJungGrabert}. The susceptibilities defined in Eqs. (\ref{chiqs}) and (\ref{chivs}) become:
\bse\bea
\chiq(t)&=&\mbox{e}^{-\gamma t/2}\Bigg(\!\cosh\left[\frac{\omega\,t}{2}\right]\!+\!\frac{\gamma}{\omega}\sinh\left[\frac{\omega\,t}{2}\!\right]\!\Bigg),\\
\chiv(t)&=&\frac{2}{\omega}\mbox{e}^{-\gamma t/2}\sinh\left[\frac{\omega\,t}{2}\right],\label{chivt}\\
\omega^{2}&=&\gamma^{2}-\frac{4\omega_{0}^{2}}{M}.
\eea\ese
Similarly, from Eqs. (\ref{D1clas}) and (\ref{A11}),
\bea
\Done(t)_{_{\mbox{\tiny{\bf{CL}}}}}&=&\frac{2\kb T}{M\omega^{2}}\mbox{e}^{-\gamma t}\Big[\cosh\big[\omega t\big]-1\Big],\nonumber\\
\sigmaone\!(t)_{_{\mbox{\tiny{\bf{CL}}}}}&=&\frac{\kb T}{\omega_{0}^{2}}\Bigg[1-\mbox{e}^{-\gamma t}
\Bigg(\frac{2\gamma^{2}}{\omega^{2}}\sinh^{2}\Big[\frac{\omega t}{2}\Big]\nonumber\\
&+&\frac{\gamma}{\omega}\sinh\big[\omega t\big]+1\Bigg)\Bigg],\label{A11pj}
\eea
Finally, $\Omega(t)$ and the diffusion constant $\Dclas(t)$ are calculated from Eqs. (\ref{Omega1}) and (\ref{Dadelman}), respectively:
\bea
\Omega(t)\!&=&\!\!\frac{2\omega_{0}^{2}}{M\omega}\sinh\Big[\frac{\omega t}{2}\Big]\!\Bigg[\cosh\Big[\frac{\omega t}{2}\Big]\!+\!\frac{\gamma}{\omega}\sinh\Big[\frac{\omega t}{2}\Big]\!\Bigg]^{\mbox{\tiny{\!\bf{-1}}}},\nonumber\\
\Dclas(t)\!&=&\!\!\frac{4\kb T}{M\gamma}\!\sinh\Big[\frac{\omega t}{2}\Big]\!
\Bigg[\sinh\!\!\Big[\frac{\omega t}{2}\Big]\!+\!\frac{\omega}{\gamma}\cosh\!\!\Big[\frac{\omega t}{2}\Big]\!\Bigg]^{\mbox{\tiny{\!\bf{-1}}}},\label{Dclas}\nonumber
\eea
The FPE, Eq. (\ref{adelman1}), reduces simply to:
\be
\frac{\partial \pcla\!(q,t,\qzero)}{\partial t}=-\Omega(t)\frac{\partial }{\partial q}q\,\pcla
+\frac{1}{2}\Dclas(t)\frac{\partial^{2}\pcla}{\partial q^{2}},\label{Prob5}
\ee
Its solution has the form of Eq. (\ref{PJprob1}) with a standard deviation obtained from Eq. (\ref{SigmaAveraged}) as follows:
\bse\bea 
\pcla\!(q,t\,|\,\qzero)&=&\!\frac{1}{\sqrt{2\pi\sigmac\!(t)}}
\!\exp\!\!\left[-\frac{(q\!-\!\chiq(t)\qzero)^{2}}{2\,\sigmac\!(t)}\right]\!\!,\label{PJprob}\\
\sigmac\!(t)&=&\frac{\kb T}{\omega_{_{0}}^{2}}\Bigg[1-\mbox{e}^{-\gamma t}\Bigg(\cosh\left[\frac{\omega\,t}{2}\right]\nonumber\\
&+&\frac{\gamma}{\omega}\sinh\left[\frac{\omega\,t}{2}\right]\Bigg)^{\!\!2}\,\Bigg].\label{sigmaPJ}
\eea\ese

Equations (\ref{PJprob}) and (\ref{sigmaPJ}) reproduce the classical results of Chandrasekha \cite{Chandrasekhar}  and AG \cite{AdelmanGarrison} if the time dependent friction kernel is substituted in the latter by  $2\gamma\delta(t)$, i.e. the Markovian limit. 
It has the structure of the Kra\-mers e\-qua\-tion\cite{Risken,Gardiner1,VanKampen2}  from which all statistical properties of the system can be derived from. 
 
 Of course, anyone would think in the alternate way to construct the FPE by reversing the method of the derivation
 of Eq. (\ref{simple}). Although it sounds logical, it gives unphysical results. The diffusion coefficient exhibits  a maximum and vanishes at long times. 

Since SJG arrives to  the Markovian version of Adelman's equation \cite{Adelman1}, then the probability density of the two approaches agrees. To demonstrate the equivalence between SJG and this work, let Adelman's version  of the bi-variate Gaussian probability density be chosen to show that effectively, $p(q,t|\qzero)$ is the contracted version of the phase space density used by SJG, i.e.
 \bea
p(q,v,t,\qzero,\vzero)&=&\frac{1}{2\pi}\frac{1}{\sqrt{\mbox{det}\mathbf{A}}}\!\exp\Big[\!\!-\!\frac{1}{2}\mathbf{y}^{\!\dagger}\!\cdot\mathbf{A}^{^{\!\!\mbox{\tiny{\bf{-1}}}}}\!\!\cdot\mathbf{y}\Big],\label{adelmanprob}
\eea
where the vector $\mathbf{y}(t)$ defines the deviation of the fluctuations of the phase point, and $\mbox{det}\mathbf{A}$ is the determinant of the matrix of their second moments  defined in Ref. \cite{Adelman1}, respectively, i.e.,
\bea
\mathbf{y}&=&
\left(\begin{array}{c}
q(t)-(\chiq(t)\qzero+\chiv(t)\vzero)\nonumber\\
v(t)-(\dot{\chiq}(t)\qzero+\dot{\chiv}(t)\vzero)\end{array}\right),\\ \nonumber \\ \nonumber
\mathbf{A}&=&\Bigg\{\begin{array}{cc}
\left<y_{_{\mbox{\tiny{\!\bf{\,1}}}}}^{2}(t)\right> & \left<y_{\mbox{\tiny{\!\bf{\,1}}}}(t)y_{\mbox{\tiny{\!\bf{\,2}}}}(t)\right> \\ 
\left<y_{\mbox{\tiny{\!\bf{\,2}}}}(t)y_{\mbox{\tiny{\!\bf{\,1}}}}(t)\right> & \left<y_{\mbox{\tiny{\!\bf{\,2}}}}^{2}(t)\right>\end{array}\Bigg\}.\nonumber
 \eea
The marginal distribution $p(q,t,\qzero)$ is found from:
 \be
 p(q,t,\qzero)=\int_{-\infty}^{\infty}\!\!\!\!d\vzero\,p_{_{\mbox{\tiny{st}}}}\!(\vzero)
 \int_{-\infty}^{\infty}\!\! \!\!dv\,p(q,v,t,\qzero,\vzero),\nonumber
 \ee
where $p_{_{\mbox{\tiny{st}}}}\!(\vzero)$ is the initial canonical velocity distribution. As a matter of fact, after solving the Gaussian integrals, the result is exactly the conditional probability density $\pcla(q,t\,|\,\qzero)$ given by Eq. (\ref{Prob5}). In AG\cite{AdelmanGarrison}, the inner integral was replaced by $p(q,t,\qzero,\vzero)$ once it was assumed to be a Gaussian centered around $\overline{q}(t)$ with standard deviation equal to Eq. (\ref{A11pj}) in the Markovian limit. 
Accordingly, the phenomenological approach worked upon in this article for the classical limit of the reduced Wigner function and the Markovian limit of AG \cite{AdelmanGarrison} are equivalents. The same argument holds the SJG method and that of Adelman\cite{Adelman1} for the whole phase space in the Ohmic regime. Therefore, the QMEs of this article and SJG describe the same phenomenon. The Wigner function of the latter is the contracted version of the former.

Finally, it has to be pointed out that AG procedure is different from this work. While in the former the FPE equation is built by assuming its solution, here it is employed as the inverse procedure: the FPE is built first and it is shown that the solution is the Gaussian of AG. The two approaches are total equivalents.
%---------------------------------------------------------------
%---------------------------------------------------------------
\subsection{Alternate picture of the classical dynamics.}
%---------------------------------------------------------------
%---------------------------------------------------------------
 An additional result is obtained by appealing to Ito's formula\cite{Gardiner1} about the relationship between SDEs and their associated FPEs. That is, the SDE whose FPE is given by Eq. (\ref{Prob5}), will be:
\bea
\dot{q}(t)&=&q(t)\,\Omega(t)+\sqrt{D_{_{\mbox{\tiny{\!cl}}}}(t)}\,dB(t),
\eea
where $B(t)$ is a Wiener process. This SDE suggests  that the problem could be analyzed as a Markovian- Brownian particle moving in the fluid with a position-time-dependent drift velocity $q(t)\Omega(t)$ with noise intensity given in terms of a  time-dependent diffusion constant. It is stochastically equivalent with the underdamped Markovian Langevin version. Even though they will sure have different outcomes, their FPEs will provide the same statistics. It could also have benefits from a mathematical point of view: its form is that of the ubiquitous SDE studied extensively in the theory of stochastic processes\cite{HanggiThomas} rather than the noise functional Eq. (\ref{qprime}). It is simpler to manipulate mathematically because of the additive white noise.

%----------------------------------------------------
%----------------------------------------------------
\section{Concluding remarks.}
%-----------------------------------------------------
%-----------------------------------------------------
The results are interesting. First, the classical limit of the conditional probability density $p(q,t|\qzero)$ matches the results  of Chandrasekhar\cite{Chandrasekhar} and Adelman and Garrison\cite{AdelmanGarrison} such as the agreement between 
Schramm {\it et al.} \cite{SchrammJungGrabert}  and the Adelman equation\cite{Adelman1} for the entire phase space. It is an indication that the FPE for the reduced Wigner function should have been correctly derived. Such a conjecture was proved by contracting the whole probability density of the phase space of Schramm {\it et al.} to a coordinate-dependent distribution. 

It is to be noted that quantum contribution in $p(q,t|\qzero)$ due to the initial correlation shows up as an additive function in the diffusion term of the  FPE, while in SJG it is spread out through the different functions that conform it. 

Since the transformation rules of the Wigner function in terms of the density operator are unique, then the proposed QME, Eq. (\ref{adjointpj}), and that of SJG, Eq. (\ref{SJG}), describes the same dynamics. In other words, the exact master equation of  SJG can be rewritten in a simpler form, by searching the evolution equation of the Wigner function regardless of the momentum. 

On the other hand, the agreement mentioned above, allows the inter\-pre\-ta\-tion of the original dynamics as a first order SDE.  It preserves all the statistical attributes of the original Langevin  equation.

A potential application of this work appeared very recently in the literature. Is the analysis by Maggaz\`{u} {\it et al.} \cite{MagazzuTalknerHanggi} on the  quantum Zeno effect, which is related to the response of a quantum system after a measurement protocol monitors its state. This could be the entry door to extend to the quantum regime the already-mentioned works\cite{SeifertAbreu,GomezSchmiedlSeifert,OscarErnestoPedro} and,  investigate how a quantum measurement affects the thermodynamical outcomes.%-----------------------------------------------------
%-----------------------------------------------------
\begin{acknowledgments}
The author thanks Prof. Nelson Pantoja for his support and valuable discussions.
\end{acknowledgments}
%---------------------------------------------------------
%---------------------------------------------------------
\appendix*
\section{Derivation of Eq. (\ref{adelman1}).}
%---------------------------------------------------------
%---------------------------------------------------------

According AG\cite{AdelmanGarrison}, the probability density is given by the Gaussian
\be
p(x,t|x_{0})=\frac{1}{\sqrt{2\pi\sigma}}\exp\Bigg(\!\!-\frac{(x-x_{0}g)^{2}}{2\sigma}\Bigg), \nonumber
\ee
where $\sigma$ and $g$ are functions of $t$. The following expressions arise:
\bea
\frac{\partial p}{\partial t}=&-&\!\!\frac{(x_{0}g\!-\!x)}{\sigma}
\Bigg(\!\!x_{0}\dot{g}\!-\!\frac{(x_{0}g\!-\!x)}{2\sigma}\dot{\sigma}\!\!\Bigg)p-\!\frac{\dot{\sigma}}{2\sigma}p  ,\label{A1}\\
\frac{\partial p}{\partial x}=&-&\frac{(x-x_{0}g)}{\sigma}p,\label{A2}\\
\frac{\partial^{2}p}{\partial x^{2}}=&-&\frac{p}{\sigma}+\Bigg(\frac{x-x_{0}g}{\sigma}\Bigg)^{\!\!2}p,\label{A3}\\
&g^{2}&\frac{d}{dt}(g^{-2}\sigma)=-2\sigma\Bigg(\frac{\dot{g}}{g}-\frac{\dot{\sigma}}{2\sigma}\Bigg).\label{A4}
\eea
From (\ref{A4}),
\be
x_{0}\dot{g}-\frac{(x_{0}g-x)}{2\sigma}\dot{\sigma}=\frac{(x-x_{0}g)}{2\sigma}g^{2}\frac{d}{dt}
(g^{-2}\sigma)+x\frac{\dot{g}}{g}.\label{A5}
\ee
Substituting (\ref{A5}) into (\ref{A1}):
\be
\frac{\partial p}{\partial t}\!=\!\Bigg[\frac{(x-x_{0}g)x\dot{g}}{\sigma g}-\frac{\dot{\sigma}}{2\sigma}\Bigg]p
+\frac{1}{2}\Bigg(\frac{x-x_{0}g}{\sigma}\Bigg)^{\!\!2}\!pg^{2}\frac{d}{dt}(g^{-2}\sigma).\nonumber\label{A5a}
\ee
Making use of (\ref{A3}),
\bea
 \frac{\partial p}{\partial t}&=&\frac{1}{2}\Bigg[g^{2}\frac{d}{dt}(g^{-2}\sigma)\Bigg]\frac{\partial^{2}p}{\partial x^{2}}\nonumber\\
 &+&\Bigg[\frac{(x\!-\!x_{0}g)x\dot{g}}{\sigma g}-\frac{\dot{\sigma}}{2\sigma}+\frac{g^{2}}{2\sigma}\frac{d}{dt}(g^{-2}\sigma)\Bigg]p,\nonumber 
\eea
 and replacing (\ref{A4}),
\be
\frac{\partial p}{\partial t}\!=\!\frac{\dot{g}}{g}\Bigg[\frac{x(x-x_{0}g)}{\sigma}-1\Bigg]p+\frac{1}{2}\Bigg[\dot{\sigma}-2\sigma\frac{\dot{g}}{g}\Bigg]\frac{\partial^{2}p}{\partial x^{2}}.\nonumber\label{A7}
\ee
From (\ref{A2}),
\be
\frac{\partial}{\partial x}(x\,p)=-\Bigg[\frac{x(x-x_{0}g)}{\sigma}-1\Bigg]p.\nonumber\label{A8}
\ee
Hence, the desired equation is:
 \bea
 \frac{\partial p}{\partial t}&=&-\frac{\dot{g}}{g}\frac{\partial}{\partial x}\big[x\,p\big]
 +\frac{1}{2}\Bigg[\dot{\sigma}-2\sigma\frac{\dot{g}}{g}\Bigg]\frac{\partial^{2}p}{\partial x^{2}}.\nonumber
 \eea
Equation (\ref{adelman1}) is obtained by identifying $p$, $g$ and the term into the brackets with  $p(q,t,\qzero)$, 
$\chiq(t)$ and $\DQ(t)$, respectively. Both, $\sigma$ and $x$ should be respectively replaced by $\sigmaQ(t)$ and $q$.

%-----------------------------------------------------
%-----------------------------------------------------
\bibliographystyle{apsrev4-1}
\bibliography{references}
%-----------------------------------------------------
%-----------------------------------------------------

\end{document}